\documentclass[12pt]{article}

\usepackage{amsmath,amsfonts,amssymb}
\usepackage{graphicx}
\usepackage{enumerate}

\makeatletter \@addtoreset{equation}{section}

\makeatletter\renewcommand\section{\@startsection {section}{1}{\z@}%
                                   {-3.5ex \@plus -1ex \@minus -.2ex}%nn
                                   {2.3ex \@plus.2ex}%
                                   {\normalfont\large\bfseries}}
\renewcommand\subsection{\@startsection{subsection}{2}{\z@}%
                                     {-3.25ex\@plus -1ex \@minus -.2ex}%
                                     {1.5ex \@plus .2ex}%
                                     {\normalfont\bfseries}}

\newcommand{\be}{\begin{equation}}
\newcommand{\ee}{\end{equation}}
\newcommand{\beq}{\begin{eqnarray}}
\newcommand{\eeq}{\end{eqnarray}}

\def\[{\left [}
\def\]{\right ]}
\def\({\left (}
\def\){\right )}

\def\CN{{\cal N}}

\def\r2{\sqrt{2}}

\newcommand{\RR}{\mathbb{R}}
\newcommand{\SL}{\mathrm{SL}}

\def\CD{{\cal D}}

\def\CF{{\cal F}}
\def\CG{{\cal G}}
\def\CH{{\cal H}}

\def\CL{{\cal L}}

\def\CN{{\cal N}}
\def\CO{{\cal O}}

\def\CS{{\cal S}}

\newcommand{\bbibitem}[1]{\bibitem{#1}\marginpar{#1}}

\def\Label#1{\label{#1}%
  \smash{\hbox to0pt{\raise1ex\hbox{\tiny[#1]}\hss}}}
\def\noLabels{\let\Label=\label}
\def\nobbibitem{\let\bbibitem=\bibitem}

\begin{document}
%\noLabels % uncomment for final production
%\nobbibitem % uncomment for final production

\begin{titlepage}

\flushright{UCB-PTH-07/21}
 \vfil\

\begin{center}

{\Large{\bf Black Holes in Supergravity: the non-BPS Branch\\ }}

\vspace{3mm}

Eric G. Gimon\footnote{e-mail: eggimon@lbl.gov}$^{\,a,b}$, Finn
Larsen\footnote{e-mail: larsenf@umich.edu}$^{\,c}$$^{,d}$ and Joan
Sim\'on\footnote{e-mail: J.Simon@ed.ac.uk}$^{\,e}$
\\

\vspace{8mm}

\bigskip\medskip
\smallskip\centerline{$^a$ \it Department of Physics, University of California, Berkeley,
CA 94720, USA.}
\medskip
\smallskip\centerline{$^a$ \it Physics Division, Lawrence Berkeley National Laboratory Berkeley, CA 94720, USA.}
\medskip
\smallskip\centerline{$^c$ \it Department of Physics, University of Michigan, Ann Arbor,
MI 48109, USA.}
\medskip
\smallskip\centerline{$^d$ \it Theory Division, CERN, CH-1211 Geneva 23, Switzerland.}
\medskip
\smallskip\centerline{$^e$ \it School of Mathematics and Maxwell Institute for Mathematical Sciences,}
\smallskip\centerline{\it King's Buildings, Edinburgh EH9 3JZ, Scotland.}

\vfil

\end{center}
\setcounter{footnote}{0}
%%%%%%%%%%%%%%%%%%%%%%%%%%%%%%%%%%%%%%%%%%%%%%%%%%%%%%%%%%%%%%%%%%%%%%%%%%%%%%%%%%%%%%%
\begin{abstract}
\noindent
%  ABSTRACT HERE
We construct extremal, spherically symmetric black hole solutions to
4D supergravity with charge assignments that preclude
BPS-saturation. In particular, we determine the ground state energy
as a function of charges and moduli. We find that the mass of the
non-BPS black hole remains that of a marginal bound state of four
basic constituents throughout the entire moduli space and that there
is always a non-zero gap above the BPS bound.

\end{abstract}
%%%%%%%%%%%%%%%%%%%%%%%%%%%%%%%%%%%%%%%%%%%%%%%%%%%%%%%%%%%%%%%%%%%%%%%%%%%%%%%%%%%%%%%%%
\vspace{0.5in}

\end{titlepage}
\renewcommand{\baselinestretch}{1.05}  %Line spacing
%%%%%%%%%%%%%%%%%%%%%%%%%%%%%%%%%%%%%%%%%%%%%%%%%%%%%%%%%%%%%%%%%%%%%%%%%%%%%%%%
%%%%%%%%%%%%%%%%%%%%%%%%%%%%%%%%%%%%%%%%%%%%%%%%%%%%%%%%%%%%%%%%%%%%%%%%%%%%%%%%%%%%%%%%%%%
%\tableofcontents

\newpage
\tableofcontents

\section{Introduction}
The construction of regular black hole solutions in supergravity has been a major research
area for many years. This effort has given a very complete understanding of the BPS black
holes and their non-extremal generalizations. However, there are many assignments of
asymptotic charges which do not correspond to regular, spherically symmetric BPS
black holes. The black holes describing such configurations are qualitatively different from
their BPS relatives. This makes them interesting, but also more complicated than the
BPS solutions. In fact, there are many simple cases where the black hole solutions have
not even been constructed. This paper seeks to fill this gap.

The shortcoming of the standard inventory of solutions can be put in context by a
well-known example. Consider the D0/D4 black hole solution with asymptotic moduli
taking canonical values. In this case the BPS black holes correspond to one sign of the
D0-brane charge (in our conventions $Q_0>0$) while the non-BPS solutions correspond to the
other sign (alas, $Q_0<0$). Moreover, as long as we consider the simplest assigment
of moduli, the BPS and non-BPS solution are related by analytical continuation: simply
invert the sign of the gauge field coupling to D0-brane charge, keeping the geometry
and all scalar fields invariant.

The point we wish to make is that this simple D0/D4 example is non-generic. In a more
general situation there are further charges present in the configuration or, equivalently,
nontrivial background moduli have been turned on. Either way, it is no longer possible to
continue analytically between the BPS and the non-BPS solutions. In fact, the two types of
solutions depend on charges so differently that their relation is nonanalytic. In this
sense the non-BPS class of solutions are reminiscent of a different phase or, at least,
a different branch of configuration space.

We focus on $N=8$ supergravity for definiteness and consider a type IIA duality frame
where all charges correspond to D-branes\footnote{Most results apply to $N=2$ and
$N=4$ supergravity after obvious changes of notation.}.
The general distinction between the two branches is encoded in the quartic
invariant which in the present context can be written as
\cite{kalloshkol,Andrianopoli:1997wi} ($\Sigma = 0..3, i = 1..3$):
\be
I_4 =  4Q_0 P^1 P^2 P^3 - 4P^0 Q_1 Q_2 Q_3 - \left( \sum_\Sigma P^\Sigma
Q_\Sigma\right)^2 + 4\sum_{i<j} P^i Q_i P^j Q_j~.
\ee
The charge configurations with $I_4 > 0$ have a BPS limit whereas those with
$I_4<0$ do not. Thus the configuration space of non-BPS solutions is as large
as that of the BPS solutions, in that they have the same number of continuous
parameters. The entropy of the black holes on either branch is given by
\cite{Larsen:1999pp}:
\be
S ={\pi\over G_N} \sqrt{ |I_4| }~.
\ee
The entropies of the two branches are therefore related in a simple way. However,
the solutions have no simple relation.

The most general spherically symmetric black hole solution in $N=8$ (or $N=4$)
supergravity can be generated by acting with dualities on a seed solution with
at least five charges \cite{Cvetic:1995uj,Cvetic:1996zq}. Five parameter
generating solutions were constructed in the BPS-case long time ago
\cite{Sen:1994eb,Cvetic:1995kv,trigiante1, trigiante2}, but on the non-BPS branch only four 
parameter solutions have been constructed so far\footnote{To our knowledge, the first
examples of non-BPS extremal solutions were found in \cite{tomas1, tomas2}.}.
The solutions we construct are the general seed solutions.

The charge assignments we focus on do in fact permit BPS solutions, at least in
some cases. Those BPS solutions are the multicenter solutions, which have been
the subject of much recent interest  \cite{4dmicro,andydecon}. These
multicenter solutions have the same asymptotic charges as those we construct,
but only exist for a limited set of possible background moduli. Nevertheless,
since the multicenter solutions are BPS, and so have less energy than the
solutions constructed here, we suspect that there is an interesting interplay
between the two classes of solutions. We hope to return to this point
elsewhere.

In this work we focus on the extremal case for conceptual clarity, but the
non-BPS branch of solutions include generalizations of these with more energy,
and with angular momentum. The solutions we construct represent
ground states, since they are extremal. From this perspective the
interesting output of the solution is the non-BPS mass formula. In the D0/D6
duality frame the mass takes the form
\cite{Rasheed:1995zv,Dhar:1998ip,Larsen:1999pp}:
$$
M  = {1\over \sqrt{8}G_4} \left( Q^{2/3} + P^{2/3} \right)^{3/2}~.
$$
We determine the generalization of this formula that includes B-fields on the
world-volume of the D6. More generally, we suspect that the non-BPS formula
reflects interesting and rather generic data that probes supersymmetry breaking
in the gravitational sector.

This paper is organized as follows.  Section 2 gives an overview of
our conventions, the equations that need to be solved and of the
various types of attractors.  Section 3 is a review of known results
for BPS attractor flows and uses a D0-D4-D4-D4 charge vector.
Section 4 gives new results for the non-BPS extremal attractor flows
in terms of one seed solution with $\overline{\rm D0}$-D4-D4-D4
charge, with some important group theory features. Section 5
dualizes these results to get a $D6-D0$ non-BPS attractor flow.
Finally, section 6 closes with a brief discussion.  The details of
our non-BPS solution are derived in an appendix.

While this paper was in preparation, some overlapping and
complementary results appeared in \cite{Gaiotto:2007ag}.

\section{The Setting}
We want to be specific about our notation and so we begin with a small review of our
setting.

\subsection{The Theory}
We work in the framework of $N=2$ supergravity coupled to a number
of vector multiplets. The bosonic action terms in the action are \cite{Ceresole:1995jg}:
\be
\CS = {1\over 8\pi G_N} \int d^4x \CL = {1\over 8\pi G_N} \int d^4x \Big[-
{R\over 2} + G_{a\bar b} \partial_\mu z^a \partial^\nu \bar z^b + \textrm{Im}\,
\big(\CN_{\Lambda\Sigma} \CF^{-\Lambda}_{\mu\nu}
\CF_{\mu\nu}^{-\Sigma}\big)\Big]~,
\ee
where $\CF^{\pm\Lambda}_{\mu\nu} = \CF^{\Lambda}_{\mu\nu} \pm
{i\over 2} \varepsilon_{\mu\nu\rho\sigma}\CF^{\Lambda\,\mu\nu}$.

We focus on the $\CN=2$ theory known as the STU-model
\cite{Cremmer:1984hj,STU,Behrndt:1996hu}. We will interpret the
model in terms of $IIA$ theory on a $T^6$ of the form $T^2\times
T^2\times T^2$. The D0/D2/D4/D6-branes wrapping the various $T^2$'s
give four magnetic and four electric charges.  The STU-model in
$\CN=2$ theory captures the essential features of extremal black
holes in the $\CN=4,8$ theories and many of it's features also
generalize well to extremal black holes in other $\CN=2$ theories
(such as those from CY compactifications).

In $\CN=2$ theory it is convenient to use the language of special geometry.  In
the STU model the prepotential and its derivative are:
\be
F = {X^1X^2X^3\over X^0}~,\qquad F_\Sigma = {\partial F\over \partial X^\Sigma}~.
\ee
We gauge fix the projective coordinates $X^\Lambda$ ($\Lambda=0,1,2,3$) so $X^0=1$
and then write $X^i = z^i = x^i - i y^i$ ($i=1,2,3$).\footnote{Some authors use
$z^i = x^i + i y^i$. Then in order to keep the K\"{a}hler metric positive, the sign
of the pre-potential $F$ is opposite. The resulting scalars are the complex conjugate of ours,
as is the central charge. The sign of the electric charges $Q_\Sigma$ are the opposite. }
The K\"{a}hler potential is:
\be
K = - \log i(\bar F_\Sigma X^\Sigma -  F_\Sigma \bar X^\Sigma) = - \ln( 8
y_1y_2y_3)~.
\ee
The corresponding metric and connection on moduli space are:
\be
G_{i\bar j} = \partial_i\partial_{\bar j}\,K = {\delta_{i\bar j}\over (2
y^i)^2}~, \qquad \Gamma^{i}_{ii} = {i\over y^i}~.
\ee
Here $i$ is not summed over. The central charge of the $\CN=2$ superalgebra
is written in terms of the superpotential $W$, as:
\be
Z = e^{K/2}\,W = e^{K/2}\,[X^\Lambda Q_\Lambda - F_\Lambda P^\Lambda]~.
\ee

\subsection{Charges}
The electric and magnetic charges are defined as:
\be
P^\Lambda = {1\over 4\pi}
\int_{S^2_\infty} \CF^\Lambda~, \qquad Q_\Sigma =  {1\over 4\pi}
\int_{S^2_\infty} \CG_\Sigma~,
\label{eqn:physchar}
\ee
where the symplectic dual field strength is:
\be
\CG_{\pm\Lambda}^{\mu\nu} = -i {\delta \CL\over \delta
\CF^{\pm\Lambda}_{\mu\nu}} = \overline\CN_{\Lambda\Sigma} \CF^{+\Sigma\,\mu\nu}~.
\ee
The physical charges (\ref{eqn:physchar}) are organized in symplectic
pairs:
\be
\Gamma \equiv (P^\Lambda,Q_\Sigma)~.
\ee
They have units of length and are related to dimensionless
quantized charges by some dressing factors. We will normalize the
asymptotic volume moduli so $y^i|_\infty = 1$ but keep the asymptotic B-fields
$x^i_\infty = B^i = {1\over V_i}\int_{V_i} B $ as free
variables\footnote{We can change to conventions where $y^i|_\infty =  v_i$
is nontrivial by taking $\tilde z^i = z^i v_i$.  The effective potential
(introduced below) satisfies:
\be
V_{BH}(P^\Lambda,Q_\Sigma, z^i) =  G_N  \tilde V_{BH}(p^\Lambda,q_\Sigma,
\tilde z^i)~,
\ee
so it is natural to associate the dressed charges $P^\Lambda,Q_\Sigma$ with
the unit normalized $z^i$'s and the quantized charges $p^\Lambda,q_\Sigma$
with volume normalized $\tilde z^i$'s.
}. Then the
dressing factors are just numerical factors
\be
P^\Lambda = C^\Lambda\,p^\Lambda~,\qquad Q_\Sigma = C_\Sigma \,q_\Sigma~,
\ee
which are essentially the masses of the underlying branes:
\beq
&&
C^0 = 2^{3/2}G_N M_{D6} = \sqrt{G_N v_6}~,\qquad
C^i = 2^{3/2} G_N M_{D4} =\sqrt{G_Nv_6}\cdot{1\over v_i}~, \\
&&
C_0 =2^{3/2} G_N M_{D0} = \sqrt{G_N\over v_6}~,
\qquad C_i = 2^{3/2} G_N M_{D2} = \sqrt{G_N\over v_6}
\cdot v_i ~. \nonumber
\eeq
Here $v_i$ are the volumes of the $T^2$'s measured in string units $v_i=V_i/(2\pi l_s)^2$.
The overall compactification volume is $v_6=v_1 v_2 v_3$ and the $D=4$
Newton's constant $G_N = l_s^2g_s^2/8v_6$.

\subsection{The Equations of Motion}
For the spherically symmetric, extremal solutions we are interested in, the metric takes the form:
\be
ds^2 = -e^{2U(\tau)} dt^2 + e^{-2U(\tau)} d\vec{x}^2~,
\ee
where the warp factor is a function of $\tau = 1/|\vec{x}|$ only.  The
functional form of the gauge fields is fixed in terms of this warp factor and
the charges and generates an effective potential for the scalars of the form \cite{Ferrara:1997tw,Levay:2007nm}:
\beq
V_{BH} &=& |Z|^2 + \sum_i |\CD_i Z|^2 \\
\nonumber &=& e^K\( |W(z_1,z_2,z_3)|^2 + |W(\bar z_1,z_2,z_3)|^2 + |W(z_1,\bar
z_2,z_3)|^2 + |W(z_1,z_2,\bar z_3)|^2\) ~.
\label{eqn:veff}
\eeq
Spherically symmetric solutions extremize the Lagrangian of the equivalent
mechanics problem:
\be
\CL_{\rm eff} = (\dot{U})^2 + G_{i\bar j} \dot{z}^i \dot{z}^{\bar j} +
e^{2U}V_{BH}~,
\label{eqn:efflag}
\ee
which amounts to solving the Euler-Lagrange equations:
\be
\ddot{U} =  e^{2U}V_{BH}~,\qquad \ddot{z}^i + \Gamma^i_{jk} \dot{z}^j \dot{z}^k
= {e^{2U}} \partial^i V_{BH}~.
\ee
Solutions must also satisfy the Hamiltonian constraint:
\be
\dot{U}^2 + G_{i\bar j} \dot{z}^i \dot{z}^{\bar j} -  e^{2U}V_{BH} = 0~.
\label{eqn:constraint}
\ee
In all these equations dots denote derivatives with respect to $\tau$. In the
appendix we make these equations explicit for the STU-model.

\subsection{Attractors}
Black hole solutions are characterized by their conserved charges. In our setting the
asymptotic data is just the charge vector $\Gamma=(P^\Lambda,Q_\Sigma)$ because
we assume spherical symmetry (so angular momentum vanishes) and extremality
(so the mass is determined by the charge as the minimal one giving a regular black hole).

In $\CN = 8$ theory there are qualitatively different classes of black hole solutions,
classified by the quartic invariant $I_4(\Gamma)$: if $I_4(\Gamma) > 0$ the solution is
BPS but if $I_4(\Gamma) < 0$ the single center solution cannot be BPS. If the
invariant is null, the solution is BPS but preserves more than
the minimum $1/8$ SUSY \cite{ferraramald,Ferrara:2007pc}.

The $\CN=2$ STU-theory inherits the quartic invariant from the $\CN=8$ theory:
\beq
\label{entropy}
I_4(\Gamma) &=& 4Q_0P^1P^2P^3 - 4P^0Q^1Q^2Q^3- \(P^\Sigma Q_\Sigma\)^2  + 4\sum_{i<j} P^iQ_i P^jQ_j~.
\eeq
It may happen that some of the $I_4(\Gamma)>0$ solutions do not preserve any of
the $\CN=2$ SUSY even though they do preserve some of the $\CN=8$ SUSY (see
\cite{Ferrara:2006em,Ferrara:2007pc}). Although such solutions are non-BPS in
the $\CN=2$ theory they have essentially the same properties as the BPS
solutions. Our interest are in the soutions with $I_4(\Gamma)<0$ which are
non-BPS whether in $\CN=2$ or $\CN=8$.

The extremal black holes, be they BPS or not, all exhibit an attractor
mechanism.  One can solve for these attractors values by minimizing
$V_{BH}(\Gamma,z^i)$ as a function of the $z^i$'s with fixed $\Gamma$.  For the
BPS solutions of the STU model, the vector-multiplet moduli, $z^i$, are {\em
all} completely fixed at the horizon and given by the following expression:
\be
\label{eqn:attractor}
z^i_{\rm fix} = {P^i + i\partial_{Q_i} I_4^{1/2}(\Gamma) \over P^0 + i
\partial_{Q_0} I_4^{1/2}(\Gamma)}~.
\ee
The non-BPS attractors with $I_4 < 0 $ are qualitatively different: out of the six real moduli
in the STU model, there are only four fixed scalars and two flat directions. The expression
for the attractor values for the four fixed scalars is now somewhat more complicated due
to certain subtle phases \cite{Kallosh:2006ib}.
The appearance of flat directions is most easily appreciated in the $D0-D6$ duality frame
where the relative size of the $T^2$'s remain undetermined upon extremization of
$V_{BH}$. We will return to this point in much more detail.

The entropy (from the horizon area) of all extremal solutions, whether BPS or
not, is essentially the effective potential (\ref{eqn:veff}) evaluated at the
extremum:
\be
S = {A\over 4G_N}= {\pi\over G_N} \left.V_{\rm BH}\right|_{\rm ext} ~,
\ee
which can be shown to give
\be
S = {\pi\over G_N}\sqrt{ | I_4(\Gamma)|}~.
\label{eqn:entrop}
\ee

\section{BPS Solutions}
Before describing new extremal non-BPS solutions to the STU-model and
the $\CN=8$ theory we review the familiar BPS solutions
\cite{Cvetic:1995uj,Tseytlin:1996bh,Gauntlett:1996pb,Balasubramanian:1996rx}.

\subsection{A Simple BPS Solution: D0-D4 Without B-fields}
A good benchmark solution is the case of a $D0-D4-D4-D4$ black hole
with $Q_0> 0$ and $P^i >0$ but $P^0=Q_i=0$. This charge
configuration is BPS. Thus its attractor values are given by \eqref{eqn:attractor} which
become:
\be
z^j_{\rm fix} = -i\sqrt{\frac{2Q_0\,P^j}{s_{jkl}P^k\,P^l}}\,,
\label{eqn:attex}
\ee
where we introduced $s_{jkl} = |\epsilon_{jkl}|$.
%If we plug these values for the scalars into the expression for $V_{BH}$ we
%recover
%\be
%V_{BH}|_{\tau = \infty} = 2G_N \sqrt{q_0p^1p^2p^3}.
%\ee
These attractor values give a natural way to write down the full solution in the simple
case where the $B$-fields, encoded in the moduli $x^i$, vanish asymptotically.
One starts with four harmonic functions:
\be
H^i = {1\over\sqrt{2}} + P^i\,\tau~,\qquad H_0 = {1\over\sqrt{2}} + Q_0\,\tau~,
\ee
and then the solution to our effective Lagrangian is:
\begin{eqnarray}
\label{simplesoln}
e^{-4U} &=& 4H_0H^1H^2H^3~,\\
\quad z^j &=& -i\sqrt{\frac{2H_0\,H^j}{s_{jkl} H^k\,H^l}}\,.
\end{eqnarray}
Inspecting the limit  $\tau \to \infty$ we recover the attractor values (\ref{eqn:attex})
and the black hole entropy (\ref{eqn:entrop}).  It is also evident that for this
solution the mass formula is just the marginal sum:
\be
G_N\,M = {1\over 2\sqrt{2}}\big(Q_0 + \sum_i P^i\big) = Z|_{\tau =0}~.
\ee
This also follows from the fact that without B-fields the phases of the
individual central charges for the D0-brane and D4-branes:
 \be
 Z_{D0} = 2^{-3/2}\,Q_0~,\qquad Z_{D4^i} = 2^{-3/2}\,P^i~,
 \ee
are completely aligned (such marginal bound states in terms of more
basic constituents were initially explored in
\cite{Rahmfeld:1995fm,Duff:1996qp}).

\subsection{The Most General BPS Solution}
By expanding the framework above one can accommodate a wider range of
asymptotic data, including the appearance of B-fields and a more general charge
vector $\Gamma$ (see {\it e.g.} \cite{4dmicro,Behrndt:1997ny,denef1,denef3}).
In addition to the four harmonics $H_0, H^i$ we need four more harmonic
functions $H^0,H_i$. Defining the constant terms:
\be
\Gamma_\infty = ({\bar P^\Sigma, \bar Q_\Lambda}) = (H^\Sigma |_{\tau
=0},H_\Lambda |_{\tau =0})~,
\label{eqn:constt}
\ee
we can write the whole set of harmonic functions compactly, as
a single charge-vector valued function:
\be
\qquad \CH(\tau) = \Gamma_\infty + \Gamma\,\tau~.
\ee
The constant terms (\ref{eqn:constt}) are subject to two conditions:
\be
\label{constraints}
I_4(\Gamma_\infty) = 1~, \qquad \langle\Gamma,\Gamma_\infty\rangle = P^\Sigma\bar Q_\Sigma -
Q_\Lambda \bar P^\Lambda = 0~.
\ee
As long as these are satisfied we can completely describe any $I_4(\Gamma) > 0$
solution succinctly using a generalization of the entropy formula to an entropy
function and a generalization of the attractor equations called the {\em
stabilization} equations \cite{Behrndt:1997ny}. The warp factor is obtained by
inserting our charge-valued harmonic function into the entropy formula in
(\ref{entropy}):
\be
\label{warp}
e^{-4U(\tau)} = I_4(\CH(\tau))~.
\ee
Similarly the solution of our scalars $z^i$ is obtained by generalizing the
attractor equations to:
\be
z^i(\tau) = {H^i(\tau) + i\partial_{H_i} I_4^{1/2}(\CH(\tau)) \over H^0(\tau) + i
\partial_{H_0} I_4^{1/2}(\CH(\tau))}~.
\label{eqn:bpssoln}
\ee
It is straightforward to verify that this formalism recovers the simple solution in
(\ref{simplesoln}).

As a special case of (\ref{eqn:bpssoln}) we note that the constants in the harmonic
function, $\Gamma_\infty$, satisfy the attractor equations. This is the attractor at
infinity~\cite{FGK}, a map between the $6$ real moduli and the $8$ constants in
the harmonic functions which are subject to the $2$ constraints (\ref{constraints}).
The fact that BPS attractors completely fix the scalars is crucial for our ability to
write the full solution in the attractor form (\ref{warp}-\ref{eqn:bpssoln}).

\subsection{D0-D4 Revisited: Non-trivial B-fields}

We now apply this machinery to our D0-D4 BPS solution and use it to
include non-trivial B-fields.  Thus we introduce general constant
terms for the harmonic function $H_0, H^i$ and also allow the
harmonic functions $H^0,H_i$ to have non-zero constant terms as
well. All these parameters are fixed in part by our choice that the
asymptotic volume moduli $y^i_\infty=1$. We parameterize the
remaining freedom in terms of the asymptotic B-field densities, $B^i
= x^i_\infty$, and a phase $\alpha$ which we will explain shortly:
\beq \bar P^0 &=&  {\sin\alpha\over \sqrt{2}}~,\quad \bar P^i =
{1\over\sqrt{2}}\,\[B^i\sin\alpha + \cos\alpha\]~,
\\ \qquad \bar Q_1 &=&
-{1\over\sqrt{2}}\,\[\sin\alpha(1-B_2B_3) - \cos\alpha(B_2 + B_3)\]~,
~({\rm and~cyclic~permutations})~, \nonumber \\
\bar Q_0 &=& {1\over\sqrt{2}}\,\[(\sum_i B^i - \prod_i
B^i)\sin\alpha + (1 - \sum_{i<j} B^iB^j)\cos\alpha\]~. \nonumber
\eeq These expressions were constructed so that the first constraint
in (\ref{constraints}) $I_4(\Gamma_\infty) = 1$ is satisfied. To
satisfy the second constraint we must choose $\alpha$ so
that($s^i_{jk} = |\epsilon_{ijk}|$): \beq
\langle\Gamma,\Gamma_\infty\rangle &=& P^i \bar Q_i - Q_0 \bar P^0 \nonumber \\
&=&  {\rm Im}
\[\(Q_0 + \sum_i{P^i}\,{s_{ijk}\over 2}(1 + i B^j)(1 + i B^k)\)
{e^{-i\alpha}\over\sqrt{2}}\] \nonumber \\
&=&  2\,{\rm Im} \[Ze^{-i\alpha}\] =0~.
\eeq
In addition to specifying $\alpha$ in the solution we therefore
find the interpretation of $\alpha$:
it is the phase of the central charge
$Z$.\footnote{This is up to a shift by
$\pi$. This ambiguity is resolved by the mass formula}
If we insert our harmonics into (\ref{warp}) we get:
\beq
e^{-4U} &=& 1 + \sqrt{2} \[  (Q_0 + \sum_i P^i (1 - s_{ijk}\,{B^j B^k\over
2})\cos\alpha + \sum_i s_{ijk}B^jP^k\sin\alpha \] \,\tau \\
&& \qquad + \CO(\tau^2) + \CO(\tau^3) + 4Q_0 P^1 P^2 P^3 \tau^4 ~.
\nonumber
\eeq
This gives the correct BPS mass and black hole entropy:
\beq
M &=& G_N^{-1} |Z|  = G_N^{-1} {\rm Re}\,\big[Ze^{-i\alpha}\big] ~,\\
S &=& {2\pi\over G_N} \sqrt{Q_0P^1P^2P^3}~.
\eeq
Expanding the general stabilisation equation (\ref{eqn:bpssoln}),
the scalars take the form:
\be
z^1 = {-H_1H^1 + H_0H^0 + H_2H^2 + H_3H^3 - i e^{-2U} \over 2(H^2H^3 -
H^0H_1)}~,~~~({\rm and~cyclic~permutations})~.
\ee
The moduli exhibit the correct asymptotic behavior, namely:
\be
z^j|_{\tau =0} \to B^j - i~,\qquad z^j|_{\tau =\infty} = -i\sqrt{Q^0\,P^j\over
{1\over 2}s_{jkl}P^k\,P^l}~.
\ee
In other words, we satisfy the boundary conditions we wanted at $r = \infty$ and flow
to the previously determined attractor values at the horizon $r=0$.

\section{Non-BPS solutions: the $\overline{\rm\bf D0}$-D4 case}

We now get to the core of our results, the non-BPS black hole solutions to
the STU-model and the $\CN=8$ supergravity theory. We first briefly review
how a non-BPS solution can be constructed by analytical continuation
and then present the more general solution that cannot be obtained this way.
We then discuss the non-BPS mass formula and the action of the dualities on
our solution.

\subsection{The Simple Non-BPS Solution}
Once again we use the canonical representative, a D0-D4-D4-D4 charge vector.
In the non-BPS case we assume $Q_0<0$, $P^i>0$ so $I_4(\Gamma) < 0$.
As mentioned in the introduction, we can derive some simple solutions by analytic
continuation from the BPS case \cite{Cvetic:1995kv,Kallosh:2006ib,Emparan:2006it}.
Thus we can start from the harmonic functions:
\be
H^i = {1\over\sqrt{2}} + P^i\,\tau~,\qquad H_0 = -{1\over\sqrt{2}} + Q_0\,\tau~,
\ee
and immediately write the non-BPS solution:
\be
\label{simple}
e^{-4U} = |4H_0H^1H^2H^3|~,\quad z^i = -i\sqrt{-H_0H^i\over {1\over 2}s_{ijk}H^jH^k}~.
\ee
In particular this gives the attractor values:
\be
z^i = -i\sqrt{-Q_0P^i\over {1\over 2}s_{ijk}P^jP^j}~.
\label{simpleattractor}
%= -{i\over v_i}\sqrt{-q_0p^i\over{1\over 2}s^i_{jk}p^jp^k}~.
\ee
Our goal is to generalize the canonical non-BPS solution (\ref{simple})
to situations where the asymptotic moduli are more general and/or there are
more charges present.

\subsection{The Seed Solution: Non-BPS Black Holes with 5 Parameters}
The experience with BPS black hole solutions suggests several strategies
which all appear to encounter difficulties:
\begin{itemize}
\item
We could determine the general attractor equations, like (\ref{eqn:attractor}) for BPS,
and then try to generate the full flow from appropriate stabilizer equations,
like (\ref{eqn:bpssoln}) for BPS.  This approach was suggested for non-BPS
solutions \cite{Kallosh:2006ib}, but it does not seem to work in the
general case where an extra phase appears in the attractor equations.
This phase difficulty can be circumvented in the first order formulation for
non-BPS black holes proposed in \cite{Ceresole:2007wx}. When considering the same
superpotential as in this work, their flow equations describe the change in the phase from its
value at asymptotic infinity to its value at the horizon. When attempting to extend this
formulation for general superpotentials, the proposal in \cite{Andrianopoli:2007gt} fails
to cover this phase change.
\item
Another strategy is to exploit the uplift to five dimensions where the situation is
simpler \cite{Cardoso:2007ky}. The difficulty with this approach is the specific
assumptions for the lift: the natural {\it ansatz} assumes that the 5D geometry
is a time fibration over a Hyper-K\"{a}hler base.  This assumption does not hold
for known solutions like the D0-D6 solution in \cite{Rasheed:1995zv,Larsen:1999pp}.
However, as it turns out, for the D2-D2-D2-D6 charge vector, the extremal non-BPS
black hole solution found in \cite{Cardoso:2007ky} is U-dual to the extremal D0-D6 black
hole we will present in the next section.
\item
The solution generating technique acting on a Schwarzschild (or
Kerr) seed solution gives the most general nonextremal solution, at
least in principle. All solutions with a BPS limit were generated
this way a long time ago \cite{Sen:1994eb,Cvetic:1995kv}. The
technique works in principle for the non-BPS branch as well: it
generated the D0-D6 solution in \cite{Larsen:1999pp}. However, the
solutions derived this way are parameterized in an unilluminating
manner that has so far resisted extraction of the general non-BPS
extremal solutions.
\end{itemize}

Instead of attempting to generalize the approaches used for the BPS solutions
we find our seed solution by direct integration of the equations of motion, generalizing
another recent computation \cite{Hotta:2007wz}. The solution identified this way has
arbitrary D0 and D4-brane charges as well as equal B-field densities, $B^i = B$,
on all three $T^2$'s.

We leave the full derivation of the
 integration of the equations of motion to appendix A and present the
 final solution in its simplest form. Once again we have
four harmonic functions:
\be
H^i = {1\over\sqrt{2}} + P^i\,\tau~,\qquad H_0 = -{1\over\sqrt{2}}(1 + B^2) + Q_0\,\tau~,
\ee
with which we can write the solution as:
\be
\label{seed}
e^{-4U} = -4H_0H^1H^2H^3 - B^2~,\qquad z^i = {B - i e^{-2U}\over s_{ijk} H^j
H^k}~.
\ee
We have already mentioned that for the non-BPS solutions $Q_0<0$, $P^i>0$.
With these assignments $H_0<0$ and $H^i>0$; so the first term in the warp factor
$e^{-4U}$ is positive definite. The constant terms in $H_0, H^i$ are such that
$e^{-4U}>1$ during the entire flow $0<\tau<\infty$.

The non-BPS solution with B-field (\ref{seed}) gives the same attractor values
for the scalars and also the same black hole entropy as the simple non-BPS
solution without a B-field (\ref{simple}). These aspects are therefore reproduced
correctly by analytical continuation from the BPS-solution discussed in section
3.3. However, we emphasize again that the full radial flow cannot be obtained in this
way.

Although we have discussed our solutions in the setting of the STU-theory,
they are readily embedded also in $\CN=4$ and $\CN=8$ supergravity. In
these contexts they serve as seed solutions which generate the most general
spherically symmetric black hole solutions upon acting with
dualities. It also appears that our solutions generalize to other $\CN=2$
theories with cubic prepotential. For such theories $s_{ijk}=|\epsilon_{ijk}|$ should be replaced by
the structure constants $c_{ijk}$, and $H^1 H^2 H^3$ should be replaced by the
invariant ${1\over 6}c_{ijk}H^i H^j H^k$.

\subsection{Duality Orbits}
The solution we have given above depends on exactly 5 parameters, four
charges and a B-field, and so it is adequate to generate the most general black hole
solution. We now review how this works in principle \cite{Cvetic:1996zq}. Explicit
examples are postponed to the next section.

The theories we consider have a continuous duality group $G$ which is spontaneously
broken by the scalar fields taking values on some coset $G/H$. Starting from a seed solution with
some canonical values of the asymptotic scalars we can generate whatever more
general values of the asymptotic scalars we desire by acting with $G$. Subsequently
we can act with $H$, which leaves the scalars invariant, to bring the charges to the
values we want to realize. That {\it all} solutions are generated this way depends on the
details of the theory.

In the STU-model $G=SL(2,\RR)^3$ and $H=U(1)^3$. Starting from our
seed solution with asymptotic moduli $z^j = B-i$ for $j=1,2,3$ we
act with $SL(2,\RR)$ on each $z^j$ to realize general moduli. There
is some redundancy in this: since the seed solution already has one
explicit modulus, $B$, there is a diagonal (the same in all three
$SL(2,\RR)$'s) duality transformation that is not needed to cover
moduli space. Having transformed to the desired point in moduli
space, the next step is to realize all charge vectors without
further changing the moduli. Since the moduli actually belong to
$(SL(2,\RR)/U(1))^3$, the $U(1)^3$ leaves moduli space invariant.
These $U(1)$'s act on the {\it relative} phases of the the central
charge $Z$ and it's covariant derivatives $\CD_{\hat i}Z$ with
corresponding actions on the charge vectors $(p^\Lambda,q_\Lambda)$
($\Lambda=0,1,2,3$). In order to realize all charge configurations
we also need to act on the {\it overall} phase of the central charge
and it's covariant derivatives. This is precisely what the redundant
$(SL(2,\RR)/U(1))^3$ duality transformation can accomplish. We see
that the fifth parameter, which we parameterize as a diagonal
$B$-field, is exactly what is needed in order that the most general
solution can be generated.

Let us also consider the general $\CN=8$ theory where $G=E_{7(7)}$
and $H=SU(8)$. Here we first act with $E_{7(7)}$ on the moduli in
the seed solution, thus reaching a generic point in moduli space. We
then transform the charges with $H=SU(8)$, which leaves moduli space
invariant. To be more precise, the central charges can be organized
in an antisymmetric $8\times 8$ matrix $x^{ab} + iy_{ab}$
($a,b=1,\ldots,8$) with skew-eigenvalues $Z_\Lambda$
($\Lambda=0,1,2,3$). The $SU(8)$ duality group transforms the
central charges in the antisymmetric representation and one can show
that it generates the most general charge vector from four real
magnitudes and the overall phase of the skew-eigenvalues (left
invariant because the $SU(8)$ has unit determinant)
\cite{Cvetic:1996zq}. Again, the extra parameter we have in our
solution is equivalent to this overall phase.

We also note that virtually identical considerations apply to the $\CN=4$ theory which
has duality group $G=SO(m,n)$ and $H=SO(m)\times SO(n)$.

The duality orbits are not special to the non-BPS branch, nor to the extremal case.
We have merely reviewed why, and in what sense, seed solutions in four dimensions
must have five parameters. From this point of view our contribution is to advocate
a particular duality frame which make the seed solutions particularly simple. The
group theoretic distinction between the BPS and non-BPS branches appears when
we consider the attractor mechanism, the subject of the next subsection.

\subsection{Non-BPS Attractors with Flat Directions}

Suppose we have followed the procedure just outlined and reached our
desired duality frame, {\it i.e.} the asymptotic scalars have been
set to realize a specific vacuum, and the charge vector has been
transformed to $\gamma =(p^\Sigma, q_\Lambda)$. We then ask: what
subgroup $\hat H$ of the duality group $G$ leaves the charge vector
$\gamma$ invariant ?

We have defined $\hat H$ so that it leaves our charges invariant,
but it generally acts non-trivially on the scalars. Denoting by
$\hat h_0\subset \hat H$ the subgroup of the duality group that
leaves both moduli and charges invariant we see that the coset $\hat
H/\hat h_0$ is the nontrivial scalar manifold generated by duality
transformations that leave the charges invariant. The physical
significance of this coset is that it corresponds to new solutions
with the same charges but different asymptotic moduli. Since the
transformations act on the entire orbit these solutions can also
have different attractor values for $z^i$. The attractor values of
the scalars are therefore {\it not} uniquely determined by the
charges if the coset $\hat H/\hat h_0$ is nontrivial at the horizon.

We defined $\hat h_0$ as the elements in the duality group $G$ that
leaves  both the charge vector $\gamma$ and the moduli invariant.
The group leaving moduli invariant (but not necessarily the charge
vectors) is $H$, the maximal compact subgroup of the full duality
group $G$. Since $\hat h_0\subset H$ we find in particular that
$\hat h_0$ is compact. For typical values of the scalars, $h_0$ will
be trivial, but at the horizon $\hat h_0$ is enhanced to the maximal
compact subgroup $H_0$ of $\hat H$~\footnote{At first it seems
naively like this enhancement should always hold, but U-duality
elements acting on the charges act with a left action on the scalars
represented as right-cosets of $G$, so compact elements of $\hat H$
need not always be in $\hat h_0$.}. Therefore, if $\hat H$ is
non-compact, there will be ${\rm dim}\hat H - {\rm dim}\hat H_0$
flat directions of the black hole potential for the given charge
vector and the corresponding moduli are fixed by spontaneous
symmetry breaking rather than dynamics. This set of flat directions
is parameterized by the coset $\hat H/\hat H_0$ (see
\cite{Bellucci:2006xz,Ferrara:2007tu} for more details).

The important point we wish to make is that $\hat H/\hat H_0$ is
trivial for BPS black holes, but non-trivial on the non-BPS branch.
In other words: the attractor mechanism determines all moduli for a
BPS charge vector, but leaves flat directions if the charge vector
is non-BPS. The key distinction between BPS and non-BPS can be
appreciated by contemplating the quartic invariant $I_4(\Gamma)$
(\ref{entropy}). Only for non-BPS charge configurations
$I_4(\Gamma)<0$ is it possible to have just two non-vanishing
charges $(P^0, Q_0)$, and in this frame there are clearly
exceptional duality transformations which remain symmetries because
$(P^i, Q_i)$ vanish. These additional symmetries persist for other
non-BPS charge configurations. We will be very explicit about how
this works when we examine the D6-D0 realization of our non-BPS
solutions in the next section.

We end this discussion by reviewing the explicit expressions for the
various groups. For the STU-model the duality group is
$G=SL(2,\RR)^3$ with maximal compact subgroup $H=U(1)^3$. In this
case the non-BPS charge configurations are left invariant by $\hat
H=SO(1,1)^2$ which has only the trivial compact subgroup $\hat H_0 =
1$. Therefore $\hat H/\hat H_0=SO(1,1)^2$ parameterize two flat
directions which decouple from the attractor mechanism.

For $\CN=8$ supergravity we need $G=E_{7(7)}$, $H=SU(8)$.  BPS
charge configurations are left invariant by $\hat H=E_{6(2)}$ with
the compact subgroup $\hat H_0 = SU(2)\times SU(6)$ leaving the BPS
charge vector invariant as well. The coset $\hat H/\hat
H_0=E_{6(2)}/SU(2)\times SU(6)$ parameterizes $40$ flat directions,
from the $\CN=2$ point of view these are the decoupled
hypermultiplet scalars. Non-BPS charge configurations are left
invariant by $\hat H=E_{6(6)}$ with the maximal compact subgroup
$\hat H_0 = USp(8)$. The coset $\hat H/\hat H_0=E_{6(6)}/USp(8)$
parameterizes $42$ flat directions, which is just large enough to
contain the forty hypermultiplet scalars and the two flat directions
we see in the STU theory.

\subsection{The Non-BPS Mass Formula and more General Moduli}

Now that we understand the symmetries which allow to easily
generate other solutions from our seed solution, we would like to
see how the extremal non-BPS black hole mass compares to the BPS
bound for any moduli.

We can appreciate the differences between BPS and non-BPS extremal
black holes better by working out their masses for the seed
solution. The non-BPS mass formula is also useful in physical
applications.

Expanding the warp factor (\ref{seed}) we find the mass:
 \be 2G_N
M_{\rm Non-BPS} = {1\over\sqrt{2}}\,\big(|Q_0| + \sum_i P^i (1 +
B^2)\big)~.
 \ee
There is a simple interpretation of this expression: the mass is
just the sum of the masses of the D0 and D4-branes individually,
with the B-field taken into account for each constituent
independently. Interestingly, this indicates that the non-BPS black
hole is a {\em marginal} bound state. In the special case where all
B-fields vanish the mass formulae are related by analytical
continuation from $Q_0<0$ to $Q_0>0$. However, the more general
expressions with B-fields turned on are not related in this way.
This indicates that the physics of the two branches is qualitatively
different, in a manner reminiscent of a system with distinct phases.

It is instructive to compare our non-BPS mass formula to the BPS
bound $M_{\rm BPS}=|Z_\Gamma|/G_N$:
 \be
 2G_N \,M_{\rm BPS} =
2|Z_\Gamma| = {1\over\sqrt{2}}\Big| Q_0 +
 \sum_i P^i\,
(1 + iB)^2\Big| \label{BPSmass}~.
 \ee
If we consider the gap between the squares of the two masses, we get
that
 \be
 \Delta = 8G_N^2 \, (M^2 - M^2_{BPS}) = 4\big|Q_0\big| \sum_i P^i  > 0~.
 \ee
Thus the additional energy associated with a non-BPS state is always
strictly positive.

We have control over the general situation, with more charges and/or
moduli turned on, due to the dualities spelled out above, in section 4.3.
The masses are {\it invariant} under such transformations and so we
immediately find:
\begin{itemize}
\item
The existence of a gap between BPS and non-BPS branch holds quite generally.
\item
The mass of a non-BPS extremal bound state is always the sum of the masses of
four $1/2$-BPS (in the N=8 language) constituents. However, the quantum numbers
of these constituents will generally be complicated.
\end{itemize}

As a special instance of these considerations one might consider the ${\bar D0}-D4$
bound state and contemplate adding on general B-fields. This can be accomplished
concretely by acting with the duality group $\hat H$ which leaves charges invariant
and acts on moduli alone. The non-BPS formula for this case can deduced explicitly
this way but it does not seem to be simple, since the stabilizer group which keeps
the quantized charge invariant scales the volumes of the various $T^2$'s so that
the dressed charges vary. The physical origin of these difficulties is that, if some
of the B-fields are not equal, the constituents will not be just $\bar{\rm D0}$-branes
and D4-branes but also D6-branes with fluxes.

\section{The D0-D6 Solution with B-fields}

In this section we work out the explicit example of a non-BPS black hole
with only D6-brane and D0-brane charge but arbitrary moduli.
There are several motivations for doing this:
\begin{itemize}
\item
We would like to compare our seed solutions to previously known non-BPS
solutions \cite{Rasheed:1995zv,Larsen:1999pp}.
\item
In the D6-D0 frame the flat directions that decouple from the attractor flow
are manifest: they correspond to adjusting the volumes of the individual $T^2$'s
without adjusting the overall $T^6$ volume. By working out the duality
transformations explicitly we can exhibit the flat directions in other
non-BPS systems as well, including our original $\overline{\rm D0}$-D4 frame.
\item
The D0-D6 system has interesting physical properties
\cite{Taylor:1997ay,Dhar:1998ip,Larsen:1999pp,Larsen:1999pu,Mihailescu:2000dn,
wittenD0D6, Emparan:2006it, Emparan:2007en}. Our supergravity
solutions add new and interesting facts about this system.
\end{itemize}

Our strategy is as follows: we determine the duality transformation
relating the D6-D0 U-duality frame to the D0-D4 frame and then use this to
find the D6-D0 solution. We move in steps of increasing complexity, starting
with no B-fields on the D0/D6, then 3 identical B-fields, and finally the more
complicated case of $3$ different B-fields.
Along the way we take the opportunity to revisit the important group $\hat H$,
introduced as the stabilizer of the charge vector. We will make the group more
explicit and further explain its significance.

We will find is useful to move back and forth between two normalizations of
our scalar fields, $z^i$ and $\tilde z^i=v_iz^i$, using the former for dressed charges
and the latter for quantized charges. Duality transformations are simplest in terms of
the quantized charges $(p^\Lambda,q_{\Sigma})$ but we will revert to the use of
dressed charges when presenting our mass formulas.

\subsection{Duality Transformations}

We want to transform between the non-BPS $\overline{\rm D0}$-D4 charges
$(q_0= q, p^i)$ used hitherto and the D0-D6 charges which we denote $(q_0, p^0)$.
To so we recall that the charges of the STU-model transform in the $(2,2,2)$ of the
$\left[SL(2,\RR)\right]^3$ duality symmetry. We can make the transformation properties of the charge
vector $(p^\Lambda, q_\Sigma)$ manifest by introducing the notation $\{a_{ijk}\}$:
\beq
p^0 = a_{111} &,& q_0 = -a_{000}~, \nonumber\\
p^1 = a_{011} &,& q_1 = a_{100}~, \\
p^2 = a_{101} &,& q_2 = a_{010}~, \nonumber\\
p^3 = a_{110} &,& q_3 = a_{001}~. \nonumber
 \eeq
The duality transformations then become:
\begin{equation}
  a^\prime_{i'j'k'}  =
  \left(M_1\right)^{~i}_{i'}\,\left(M_2\right)^{~j}_{j'}\,\left(M_3\right)^{~k}_{k'}\,
  a_{ijk}~.
 \label{charget}
\end{equation}
We have introduced three independent $\SL(2,\RR)$ transformations, $M_j$
($j=1,2,3$), whose action on the complex moduli $\tilde z^j = \tilde x^j
-i\,\tilde y^j$ is:
 \begin{equation}
  M_j=
    \begin{pmatrix}
       a_j & b_j \\
       c_j & d_j
    \end{pmatrix}
  :\quad \tilde z^j \longrightarrow \frac{a_j\,\tilde z^j + b_j}{c_j\,\tilde z^j + d_j} \,.
\end{equation}
The $M_i$'s that dualize from the D6-D0 frame to the $\overline{\rm D0}$-D4 frame
must satisfy the eight relations:
\beq
-q &=& -a_1a_2a_3\, q_0 + b_1b_2b_3\, p^0~, \nonumber \\
0 &=& - a_1a_2\,c_3\, q_0 + b_1b_2\,d_3\, p^0~,({\rm and~cyclic~permutations})~,
\label{eqn:transf}
\\
p^i &=& -{1\over 2}s_{ijk}\,a_ic_jc_k\, q_0 + {1\over 2}s_{ijk}\,b_id_jd_k\,p^0 \,, \nonumber \\
0 &=& -c_1c_2c_3\, q_0 + d_1d_2d_3\, p^0~, \nonumber
\eeq
where $s_{ijk} = | \epsilon_{ijk}|$.
There are no solutions to these equations unless the product $q\,p^1\,p^2\,p^3$ is
negative, as expected because we must be in the non-BPS branch $(I_4<0)$ in order to dualize to the
D0-D6 system. The D0/D6 charges $p^0$ and $q_0$ can have any signs, which is also as
expected since $I_4 <0$ independently of those. Without loss of generality, we take $\{q<0\,,p^i>0\}$
and $\{p^0\,,q_0 >0\}$. With these assignments, the $\SL(2,\RR)$ matrices that map the
D0-D6 charge vector into the $\overline{\rm D0}$-D4 configuration are:
\begin{equation}
  M_i = \frac{-1}{\sqrt{2\lambda\,\rho_i}}\,
\begin{pmatrix}
\rho_i\lambda & -\rho_i \\
\lambda & 1
\end{pmatrix}\,,
\quad \rho_i = \sqrt{\frac{-qp^i}{{1\over 2}s_{ijk}p^jp^k}}\,, \quad
\lambda=\left(\frac{p^0}{q_0}\right)^{1/3}~.
\label{eqn:matrixtransf}
\end{equation}
The duality invariant $I_4$ is the same in either frame:
\begin{equation}
  I_4 = 4\,q\,p^1\,p^2\,p^3 = -(p^0\,q_0)^2 <0\,.
\end{equation}
This is necessary for the transformations $M_i$ to belong to $\SL(2,\RR)$  and for the consistency of the relations \eqref{eqn:transf} with the matrix \eqref{eqn:matrixtransf}. We will also need the inverse matrices, mapping the $\overline{\rm D0}$-D4 system into D0-D6:
\begin{equation}
\label{inverses}
M^{-1}_i = \frac{-1}{\sqrt{2\lambda\,\rho_i}}\,
\begin{pmatrix}
1 & \rho_i \\
-\lambda & \rho_i\,\lambda
\end{pmatrix}\,.
\end{equation}

The transformation matrix \eqref{eqn:matrixtransf} is not the most general one mapping the D0-D6 charge vector into the $\overline{\rm D0}$-D4 configuration. There exists a two parameter family of transformations by considering different $\lambda_i$ $(i=1,2,3)$ subject to the constraint
$\lambda_1\,\lambda_2\,\lambda_3 = p^0/q_0$. The existence of such general transformations agrees
with the conclusion in our previous duality group orbit discussion.

\subsection{Flat Directions Made Explicit}
We now have the explicit formulae needed to make the abstract
discussion of flat direction in the previous subsection more
explicit. By definition, the subgroup $\hat H$ is the subgroup of
$\left[\SL(2,\RR)\right]^3$ that leaves a given charge vector
$\Gamma$ invariant. This subgroup is particularly simple to
characterize for the D0-D6, since $\Gamma$ has only two
non-vanishing components.

The explicit $SL(2,\RR)^3$ transformation is (\ref{eqn:transf}), but now with the D0-D6
charges on the left hand side as well as the right hand side. The solutions of the equations
are the elements of $\hat H$. We find $\SL(2,\RR)$ matrices of the form:
\begin{equation}
N^{(60)}_i = \begin{pmatrix}
e^{\alpha_i} & 0 \\
0 & e^{-\alpha_i}
\end{pmatrix}, \qquad \sum_i \alpha_i = 0\,.
\end{equation}
Thus, in the D0-D6 frame, the action of this subgroup on the complex moduli is
equivalent to a $T^6$ volume preserving rescaling. That is, each moduli $\tilde{z}^i$
is rescaled:
\begin{equation}
  \tilde{z}^i \rightarrow e^{2\alpha_i}\,\tilde{z}^i\,,
  \label{d0d6hhat}
\end{equation}
keeping the product $\tilde{z}^1\,\tilde{z}^2\,\tilde{z}^3$ invariant since $\sum_i \alpha_i = 0$.

The $\hat{H}$ action in the $\overline{\rm D0}$-D4 frame is more complicated,
but it can be obtained by mapping the charge vector $\Gamma_{\overline{\rm
D0}-D4}$ to $\Gamma_{D0-D6}$, acting with $N^{(60)}_i$ in that frame, and mapping
the charge vector back to the original $\overline{\rm D0}$-D4 frame. In other
words, the action of $\hat{H}$ is given by the conjugated matrices
$N^{(40)}_i = M_i\cdot N^{(60)}_i\cdot M_i^{-1}$:
\begin{equation}
N^{(40)}_i =
\begin{pmatrix}
\cosh\alpha_i & \rho_i \sinh\alpha_i \\
\rho_i^{-1}\sinh\alpha_i & \cosh\alpha_i
\end{pmatrix},\qquad \sum_i \alpha_i = 0\,.
\end{equation}
One can explicitly check that the $\overline{\rm D0}$-D4 charges are left
invariant under these transformations. This action does mix the volume and
metrics, as can be seen by writing the explicit action on the
complex moduli fields $\tilde{z}^j = \tilde{x}^j -i\,\tilde{y}^j$:
\begin{equation}
  {\tilde z}^j \rightarrow \frac{\cosh 2\alpha_j\,\tilde{x}^j
  + (1/2)\,\sinh 2\alpha_j\,[\rho_j + \rho_j^{-1}\,((\tilde{x}^j)^2+(\tilde{y}^j)^2)] -i\,\tilde{y}_j}{\cosh^2\alpha_j
  + \rho_j^{-2}\,\sinh^2 \alpha_j\,((\tilde{x}^j)^2+(\tilde{y}^j)^2) + \rho_j^{-1}\,\sinh 2\alpha_j\,\tilde{x}^j}\,.
  \label{d0d4hhat}
\end{equation}
This action applies to the entire flow. For example, we can act on the simple non-BPS solution
(\ref{simple}). This will modify the simple attractor behavior (\ref{simpleattractor}) which, in particular will
include B-fields after this transformation. This is despite the fact that charges have not changed. Thus
the transformed attractor is sensitive to data beyond the charges, namely the asymptotic moduli
(which are computed by (\ref{d0d4hhat}) acting on the asymptotic moduli).

\subsection{D0-D6 with no B Fields}

Let us start with the seed solution written with undressed charges.  Define a a
new B-field, $b = B/G_N$, and the rescaled "undressed" harmonics $h_0 =
C_0^{-1} H_0$ and $h^i = (C^i)^{-1} H^i$.  Then the seed solution can be
written as:
\be
\tilde z^i = \frac{b - i \sqrt{-4h_0h^1h^2h^3 - b^2}}{s_{ijk}h^jh^k}~.
\ee
In order to have zero B-fields in the dual D0-D6 frame, we want the transformed
moduli, $\tilde\phi^i = M_i^{-1}(\tilde z^i)$, to be purely imaginary.
Inspection of \eqref{inverses} reveals this can only occur when:
\be
\label{fixedmod}
|\tilde z^i| = \rho^i\quad \forall\tau \qquad \Leftrightarrow \qquad h^i = p
\,(h + \tau)~,~~~h_0 = q\, (h + \tau)~,
\ee
for some constant $h$.  Thus, the D0-D6 scalars are:
\be
\tilde\phi^i = \lambda^{-1}\,\frac{1 + \rho_i^{-1}\tilde z^i}{1 -
\rho_i^{-1}\tilde z^i} = - i \lambda^{-1}\sqrt{|p^0q_0|\,(h+\tau)^2 + b\over
|p^0q_0|\,(h+\tau)^2 - b} = - i \lambda^{-1}\sqrt{|P^0Q_0|\,(h+\tau)^2 + B\over
|P^0Q_0|\,(h+\tau)^2 - B}~,
\label{modd0d6}
\ee
and the warp factor becomes:
\be
e^{-4U} = \sqrt{(P^0Q_0)^2(h + \tau)^4 - B^2}~.
\label{warpd0d6}
\ee
All the volumes $v_i$ are equal on the $D0-D6$ side. This feature can easily be
relaxed by acting with $\hat H$ (\ref{d0d6hhat}), which will make the $v_i$ general,
while keeping the overall volume $v_1 v_2 v_3$ fixed.

A less trivial task is to rewrite $B, h$ in terms of physical quantities in
the D0-D6 frame. First, from the normalization of the volume moduli at
infinity, we get:
\be
\Lambda^2 \equiv \({P^0\over Q_0}\)^{2/3}= \lambda^2 v_i^2  = {|P^0Q_0|\,h^2
+ B\over |P^0Q_0|\,h^2 - B}~.
\ee
Second, requiring that the warp factor (\ref{warpd0d6}) asymptotes to
Minkowski space we find:
\be
 (P^0Q_0)^2\,h^4 - B^2 = 1~.
\ee
This allows us to solve for $\{B,\,h\}$ in terms of the D0-D6 dressed charges
$\{Q_0,\,P^0\}$:
\beq
B &=& {1\over 2 }(\Lambda - \Lambda^{-1})= {1\over 2 (P^0Q_0)^{1/3}}
[(P^0)^{2/3} - (Q_0)^{2/3}]~,\\
h &=& {(\Lambda + \Lambda^{-1})^{1/2}\over \sqrt{2} (P^0Q_0)^{1/2}} =  {1\over
\sqrt{2}(P^0Q_0)^{2/3}}[(P^0)^{2/3} + (Q_0)^{2/3}]^{1/2}\,.
\eeq
With these identifications the D0-D6 solution (\ref{modd0d6}-\ref{warpd0d6})
agrees with the one that was constructed directly in \cite{Rasheed:1995zv,Larsen:1999pp}.

Expanding the warped factor at first order in $\tau$ gives
the mass of the D0-D6 system:
\be
2^{3/2} G_N M=  2^{3/2} (P^0Q_0)^2 h^3 = [(P^0)^{2/3} + (Q_0)^{2/3}]^{3/2}~.
\ee
Note that we can rewrite the mass formula as:
\be
2^{3/2} G_N M = 4 \({P^0\over4}\)[ 1 + \Lambda^{-2}]^{3/2}~,
\ee
which is exactly the sum of the masses of four D6-branes with charge $P^0/4$
and fluxes $\Lambda^{-1}$ on each $T^2$ with signs $(+++),(-++),(+-+),(++-)$ as
already seen in \cite{Larsen:1999pu,Emparan:2006it}. This is an example
of the more general phenomenon that the non-BPS mass can be interpreted
as a marginal bound state of elementary constituents.

\subsection{D0-D6 with Equal B-fields}

To turn on equal B-fields we need a unique non-vanishing real part for the
transformed $\tilde{\phi}^i$ for $i=1,2,3$. This is achieved by considering the
choice of harmonic functions:
\begin{equation}
  h^i = p\,(h + \tau)\,, \quad i=1,2,3 \quad \text{and} \quad  h_0 = q (k + \tau)\,.
\end{equation}
Using the inverse matrices \eqref{inverses} and the above form for the harmonic
functions, the transformed complex moduli $\tilde{\phi}^j = \tilde{x}^j -
i\,\tilde{y}^j$ are:
\begin{eqnarray}
   \tilde{x}^j &=& \lambda^{-1}\,\frac{(h-k)(h+\tau)}{(h+\tau)(h+k+2\tau)- 2{\bar b}}~, \\
   \tilde{y}^j &=&  2\lambda^{-1}\frac{\sqrt{(k+\tau)(h+\tau)^3-{\bar b}^2}}{(h+\tau)(h+k+2\tau) - 2{\bar b}}~,\\
\end{eqnarray}
where we introduced ${\bar b} = b/\sqrt{|I_4|}$.

Let us proceed as in the vanishing B field configurations. First, let us make
sure the metric is asymptotically Minkowski, by requiring the warped factor to
vanish at infinity:
\begin{equation}
  U(\tau=0) = 0 \quad \Leftrightarrow \quad |I_4|\,k\,h^3 - b^2 = 1\,.
 \label{cond1}
\end{equation}
Second, let us normalize the moduli at infinity:
\begin{equation}
  \lim_{\tau\to 0} z^j = B - i  \qquad\Rightarrow\qquad  B =
\sqrt{|I_4|}\,h\,\chi~,\qquad \frac{1}{2}\,\sqrt{|I_4|}\,h(h + k) - b =
\Lambda^{-1}~.
 \label{cond2}
\end{equation}
Above, we did introduce the notation $\chi = (h-k)/2$, we used \eqref{cond1}
when necessary, and we already took into account the scaling relation
$\tilde{z}^j=z^j\,v_j$.

Equations \eqref{cond1}-\eqref{cond2} provide three constraints that allow us
to fix the constants $\{k,\,h,\,b\}$ in terms of the physical parameters
$\{B,\,Q_0,\,P^0\}$. First, from the second equation in \eqref{cond2}:
\begin{equation}
 (b+ \Lambda^{-1})^2 = |I_4|\,h^2(h-\chi)^2
 = |I_4|\,h^3(h- 2\chi) + |I_4|\,h^2\chi^2 = 1 + b^2 + B^2\,,
 \end{equation}
where, in the last step, we used \eqref{cond1} and the first equation in
\eqref{cond2}. The above determines $b$:
\begin{equation}
 b = \frac{1}{2}\Big[\Lambda\,(1 +  B^2) - \Lambda^{-1}\Big]\,.
\end{equation}
Inserting this back into the second equation in \eqref{cond2} we get:
\begin{equation}
\sqrt{|I_4|}\,h^2 -  B = \Lambda^{-1} + b \qquad\Rightarrow\qquad
 \sqrt{|I_4|}\,h^2 = B + \frac{1}{2}\Big[\Lambda(1 + B^2) + \Lambda^{-1}\Big]\,.
\end{equation}
This fixes $h$, and so we can use the first equation in \eqref{cond2} to
determine $\chi$ (or $k$).

The mass can be obtained, as usual, by studying the first order asymptotic
correction to the warped factor. This gives:
\begin{multline}
  2G_N\,M = \frac{1}{2}\,|I_4|\,h^2\,\left(3k+h\right)=\sqrt{|I_4|}\,h\,\left(2\sqrt{|I_4|}\,h^2 -3\sqrt{|I_4|}\,h\,\chi\right) \\
  =  \sqrt{\frac{\sqrt{|I_4|}}{2}}\,\Big[ \Lambda(1 +  B^2) + \Lambda^{-1} + 2B\Big]^{1/2}
 \Big[ \Lambda(1 +  B^2) + \Lambda^{-1} - B\Big]\,.
\end{multline}
This can be written in terms of the dressed charges $\{Q_0,\,P^0\}$ as:
\begin{multline}
  2G_N\,M = \frac{1}{\sqrt{2}} \Big[(P^0)^{2/3}(1 +  B^2)
 + (Q_0)^{2/3} + 2B (P^0Q_0)^{1/3}\Big]^{1/2} \\
 \times\Big[(P^0)^{2/3}(1 +  B^2)
 + (Q_0)^{2/3} - B (P^0Q_0)^{1/3}\Big]\,.
\end{multline}
This formula make look a bit perplexing at first.  Once again, however, it has
a natural interpretation sum of the mass of four D6-branes with fluxes $|F_i| =
\Lambda^{-1}$ coming with signs $(+++),(-++),(+-+),(++-)$.  The mass formula
above is then re-written as:
\beq
2^{3/2} G_N M&=& {P^0\over 4}\Big(1 + (\Lambda^{-1} + B)^2\Big)^{1/2} \Big(1 +
(\Lambda^{-1} + B)^2\Big)^{1/2}\Big(1 + (\Lambda^{-1} + B)^2\Big)^{1/2} \nonumber \\
&+& {P^0\over 4}\Big(1 + (\Lambda^{-1} + B)^2\Big)^{1/2} \Big(1 + (\Lambda^{-1}
- B)^2\Big)^{1/2}\Big(1 + (\Lambda^{-1} - B)^2\Big)^{1/2} \\
&+& {P^0\over 4}\Big(1 + (\Lambda^{-1} - B)^2\Big)^{1/2} \Big(1 + (\Lambda^{-1}
+ B)^2\Big)^{1/2}\Big(1 + (\Lambda^{-1} - B)^2\Big)^{1/2} \nonumber \\
&+& {P^0\over 4}\Big(1 + (\Lambda^{-1} - B)^2\Big)^{1/2} \Big(1 + (\Lambda^{-1}
- B)^2\Big)^{1/2}\Big(1 + (\Lambda^{-1} + B)^2\Big)^{1/2} \nonumber.
\eeq
The variables $P^0$ and $\Lambda$ only depend on the product of the three $T^2$
volumes, so adjusting the volumes while keeping their product invariant will
leave our mass formula invariant, once again the action of $\hat H$.

\subsection{D0-D6 with Non-equal B-fields: the Most General Solution}

To turn on three different B fields, we need to consider the most general set of harmonic
functions:
\begin{equation}
  h^i = p^i\,(k^i + \tau)\,, \quad i=1,2,3 \quad \text{and} \quad h_0 = q\,(a+\tau)\,.
\end{equation}
Using the inverse matrices \eqref{inverses} and the $h_I$ defined above, the transformed complex moduli $\tilde{\phi}^j = \tilde{x}^j -i\,\tilde{y}^j$ are:
\begin{eqnarray}
   \tilde{x}^i &=& \lambda^{-1}\,\frac{(k^j+\tau)(k^l+\tau)-(a+\tau)(k^i+\tau)}{(k^j+\tau)(k^l+\tau) + (a+\tau)(k^i+\tau)- 2{\bar b}}\,, \label{xgenB} \\
   \tilde{y}^i &=&  2\lambda^{-1}\frac{\sqrt{(a+\tau)(k^1+\tau)(k^2+\tau)(k^3+\tau)-{\bar b}^2}}{(k^j+\tau)(k^l+\tau) + (a+\tau)(k^i+\tau)- 2{\bar b}}\,. \label{ygenB} \\
\end{eqnarray}
where we introduced ${\bar b} = b/\sqrt{|I_4|}$.

Let us map the constants in $h_I$ to the dressed charges and $B_i$ fields.
Requiring the warped factor to vanish at infinity is equivalent to:
\begin{equation}
  U(\tau=0) = 0 \quad \Leftrightarrow \quad |I_4|\,a\,k^1\,k^2\,k^3 - b^2 = 1\,.
 \label{cond1b}
\end{equation}
The normalisation of the moduli fields at infinity gives rise to:
\begin{eqnarray}
  \lim_{\tau\to 0} z^j = B_j - i  \qquad\Rightarrow\qquad  & & B_j =
\frac{1}{2}\sqrt{|I_4|}\,\left(\frac{1}{2}s_{ijl}\,k^j\,k^l-a\,k^i\right)\,,
\label{cond2b1} \\
& &   k^j\,k^l+ a\,k^i - 2{\bar b} =
2\frac{\Lambda_i^{-1}}{\sqrt{|I_4|}}\,,
 \label{cond2b2}
\end{eqnarray}
where we defined $\Lambda_i = \lambda\,v_i$. To derive these expressions, we used \eqref{cond1b} when necessary, and we already took into account the scaling relation $\tilde{z}^j=z^j\,v_j$.

Using \eqref{cond2b2}:
\begin{eqnarray}
  \left(b+\Lambda_i^{-1}\right)^2 &=& \frac{|I_4|}{2}\,\left(\frac{1}{2}s_{ijl}k^j\,k^l-a\,k^i + 2a\,k^i\right)^2 \\
  &=& B_i^2 + |I_4|\,a\,k^i\,\frac{1}{2}s_{ijl}\,k^j\,k^l = 1+b^2 + B_i^2\,,
\end{eqnarray}
where in the last step we used \eqref{cond1b}, allows us to determine $b$ :
\begin{equation}
  b = \frac{1}{2}\left(\Lambda_i\,(1+B_i^2) - \Lambda_i^{-1}\right)\,.
 \label{bmatch}
\end{equation}
Since the left hand side is a given number $(b)$, we do explicitly see that for a given value
of the $B_i$ fields, the volumes $v_i$ are entirely fixed at this point. The above equation gives
rise to two conditions:
\begin{equation}
  \Lambda_1\,(1+B_1^2) - \Lambda_1^{-1} = \Lambda_2\,(1+B_2^2) - \Lambda_2^{-1} =
  \Lambda_3\,(1+B_3^2) - \Lambda_3^{-1}\,.
\end{equation}
There exists a third one coming from the fact that:
\begin{equation}
  \Lambda_1\,\Lambda_2\,\Lambda_3 = \Lambda^3\,.
\end{equation}
These equations provide an implicit map between the "fluxes" $\Lambda_i$ and the $B_i$ fields.
A better parameterisation can be achieved by introducing a new set of parameters $\beta_i$ as follows:
\begin{equation}
  \Lambda_i = \frac{e^{\beta_i}}{\sqrt{1+B_i^2}}\,.
\end{equation}
In terms of these, the above conditions are:
\begin{eqnarray}
  \sqrt{1+B_1^2}\,\sinh\beta_1 &=& \sqrt{1+B_2^2}\,\sinh\beta_2 = \sqrt{1+B_3^2}\,\sinh\beta_3\,,
  \label{id1} \\
  \Lambda^3 &=& \frac{e^{\beta_1+\beta_2+\beta_3}}{\sqrt{1+B_1^2}\,\sqrt{1+B_2^2}\,\sqrt{1+B_3^2}}\,.
\end{eqnarray}

Plugging \eqref{bmatch} into \eqref{cond2b2} allows us to fix $a\,k^i$:
\begin{equation}
  \sqrt{|I_4|}\,a\,k^i = \frac{1}{2}\left(\Lambda_i^{-1} + \Lambda_i\,(1+B_i^2) - 2B_i\right) =
  \sqrt{1+B_i^2}\,\cosh\beta_i - B_i\,.
  \label{id4}
\end{equation}
Using this into \eqref{cond2b1} fixes $s_{ijl}k^j\,k^l$ to be:
\begin{equation}
  \sqrt{|I_4|}\,\frac{1}{2}\,s_{ijl}k^j\,k^l =
  \frac{1}{2}\left(\Lambda_i^{-1} + \Lambda_i\,(1+B_i^2) + 2B_i\right)= \sqrt{1+B_i^2}\,\cosh\beta_i + B_i\,.
\label{id2}
\end{equation}

The mass can be obtained, as usual, by studying the first order asymptotic
correction to the warped factor. This gives:
\begin{equation}
  2G_N\,M = \frac{1}{2}|I_4|\,\left(k^1\,k^2\,k^3 + a\,(k^1\,k^2 + k^1\,k^3+ k^2\,k^3)\right)\,.
\end{equation}
To write this in terms of the physical charges and B moduli, let us first multiply the three independent
equations \eqref{id2}. This allows us to determine the product $k^1\,k^2\,k^3$ :
\begin{equation}
  k^1\,k^2\,k^3 = |I_4|^{-3/4}\,\prod_{i+1}^3 \left[B_i + \sqrt{1+B_i^2}\,\cosh\beta_i\right]^{1/2}\,.
 \label{id3}
\end{equation}
If we now multiply \eqref{id4} with \eqref{id3} and divide by \eqref{id2}, we can determine $a\,k^i\,k^j$:
\begin{multline}
  a\,k^1\,k^2 =  |I_4|^{-3/4}\, \left[\sqrt{1+B_1^2}\,\cosh\beta_1- B_1\right]^{1/2}\cdot \left[B_2 + \sqrt{1+B_2^2}\,\cosh\beta_2\right]^{1/2}\, \\
  \cdot \left[\sqrt{1+B_3^2}\,\cosh\beta_3 - B_3\right]^{1/2}\,,
\end{multline}
with cyclic permutation expressions for the analogous remaining terms appearing in the mass formula.
We stress that to derive the above relation it may be convenient to use the identity:
\begin{multline}
  \left(\sqrt{1+B_i^2}\,\cosh\beta_i - B_i\right) \left(\sqrt{1+B_i^2}\,\cosh\beta_i + B_i\right) = \\
  \left(\sqrt{1+B_j^2}\,\cosh\beta_j - B_j\right) \left(\sqrt{1+B_j^2}\,\cosh\beta_j + B_j\right)~,
\end{multline}
for any pair $\{i,\,j\}=\{1,2,3\}$, which is a consequence of conditions \eqref{id1}.

To simplify the final mass formula, it is convenient to use:
\begin{equation}
  \sqrt{1+B_i^2}\,\cosh\beta_i \pm B_i = \frac{\Lambda_i}{2}\,\left[1 + (\Lambda_i^{-1} \pm B_i)^2\right]\,.
\end{equation}
This allows us to write the mass formula for arbitrary $B_i$ fields
as: \beq \hspace{-30pt}
2^{3/2}\,G_N\,M &=&  \frac{P^0}{4}\Big(1 +
(\Lambda_1^{-1} + B_1)^2\Big)^{1/2} \Big(1 +
(\Lambda_2^{-1} + B_2)^2\Big)^{1/2}\Big(1 + (\Lambda_3^{-1} + B_3)^2\Big)^{1/2} \nonumber \\
&+& \frac{P^0}{4}\Big(1 + (\Lambda_1^{-1} + B_1)^2\Big)^{1/2} \Big(1 +
(\Lambda_2^{-1} - B_2)^2\Big)^{1/2}\Big(1 + (\Lambda_3^{-1} - B_3)^2\Big)^{1/2} \\
&+& \frac{P^0}{4}\Big(1 + (\Lambda_1^{-1} - B_1)^2\Big)^{1/2} \Big(1 +
(\Lambda_2^{-1} + B_2)^2\Big)^{1/2}\Big(1 + (\Lambda_3^{-1} - B_3)^2\Big)^{1/2} \nonumber \\
&+& \frac{P^0}{4}\Big(1 + (\Lambda_1^{-1} - B_1)^2\Big)^{1/2} \Big(1 +
(\Lambda_2^{-1} - B_2)^2\Big)^{1/2}\Big(1 + (\Lambda_3^{-1} + B_3)^2\Big)^{1/2}
\nonumber.
\eeq
Once again, the mass remains marginal and can still be interpreted as the sum of the masses of four D6-branes with the
appropriate fluxes. In this case, the fluxes are determined implicitly as a function of the B fields.

As we emphasized above, the volumes of the different $T^2$'s are fixed in the current solution.
We can generate a more general solution in which these volumes are generic, but having a fixed
$T^6$ volume, as required by the attractor mechanism. This is achieved by the action of $\hat H$.
Since the mass $M$ only depends on the charges $\{P^0,\,Q_0\}$ and the three $B^i$, and all these
are left invariant by $\hat H$, the mass will remain marginal along all the moduli space in the D0-D6 system. Actually, using U-duality, we can extend this marginality claim on the   $\overline{D0}-D4$ side even for non-equal B-fields.

\section{Discussion}

In this paper we presented constructions of the extremal non-BPS black holes in the
STU-model along with their embeddings into
the $\CN =8$ and $\CN=4$ supergravity theories.
In addition to our detailed formulae, we highlight several qualitative lessons:
\begin{itemize}
\item
The extremal mass formula for the non-BPS black holes is generally {\it not} related
by analytical continuation to the BPS mass formula. For generic charge vector and/or
generic moduli the mass formula differ qualitatively between the two branches.
This may indicate that analysis of the non-BPS black holes relying on analytical
continuation is misleading.
\item
The non-BPS charge configurations have a canonical split into four subparts,
$\Gamma = \Gamma_0 +\Gamma_1 + \Gamma_2 + \Gamma_3$,
realized via U-duality to the ${\bar D0}-D4-D4-D4$ frame. In this duality
frame the non-BPS mass formula takes the form of a marginal bound state of the
subparts. This is quite different from the experience with BPS-black holes at generic
points in moduli space.
\item
The flat directions previously noticed for non-BPS attractors extend to the whole
flow. This is due to the existence of a nontrivial subgroup $\hat H$ of the duality
group $G$ which leaves non-BPS charge vectors invariant, while acting non-trivially
on the scalars in $G/H$.

\end{itemize}
We expect all these results to extend to all symmetric supergravity
theories by extension from the STU case. On the other hand, the flat
directions we see are intricately tied to the symmetries of our
scalar manifold so we don't expect to see such flat directions occur
naturally in non-symmetric $\CN=2$ theories.

The marginality property of our non-BPS mass formula begs for an explanation.
Such an explanation might possibly be found by following up on the surprising
observation in \cite{Cardoso:2007rg} that the near horizon regions for our
non-BPS black holes can be lifted to super-symmetric five dimensional solutions,
with "supersymmetry without supersymmetry" as in
\cite{Nilsson:1984bj,Duff:1997qz}. In any case, an explanation is beyond the
scope of this paper.

One might also wonder if the marginality property extends to non-BPS extremal black
holes (with or without wrapped D6-brane charge) in other $\CN=2$ theories, especially
those with cubic prepotentials. Finally, it would be interesting to know how the marginality
property survives $\alpha'$ corrections to the action.  In this context, it would be particularly
interesting to look at cases where $\Gamma$ is properly quantized
but the subparts $\Gamma_\mu$ are not.

\section*{Acknowledgements}
We would like to thank Oleg Lunin for early discussions on this
work. JS would like to thank the University of California at Berkeley for
hospitality during part of this work. The work of EG is supported in
part by the US DOE under contract No. DE-AC03-76SF00098 and the
Berkeley Center for Theoretical Physics.  The work of FL is
supported by DoE under grant DE-FG02-95ER40899. The work of JS was partially
funded by DOE under the contract number DE-AC02-05CH11231.

\newpage

\appendix

\section{Non-BPS D0-D4 with General Moduli: Derivation}
\label{solution}

We consider a non-BPS system with D0-brane charge $Q_0<0$ and three D4-brane
charges $P^i>0$ for $i=1,2,3$. In order to be sufficiently general we include
arbitrary complex moduli fields $z_j = x_j- i\,y_j$ $(j=1,2,3)$ at the outset.

The Lagrangian (\ref{eqn:efflag}) for the analogue mechanics problem is:
\begin{equation}
  {\cal L}_{\rm eff} = (U^\prime)^2 +\frac{1}{4}\sum_{i=1}^3\left[
\frac{(y^\prime_i)^2+(x^\prime_i)^2}{y^2_i}\right] + e^{2U}\,
V_{\text{BH}}~,
\end{equation}
where the potential $V_{\text{BH}}$ in the case of the STU-model is:
\hspace{-20pt}
\beq
  \hspace{-30pt} V_{\text{BH}}(x_i, y_i) &=&
\frac{Q_0^2}{2}\,\frac{1}{y_1\,y_2\,y_3} + Q_0\,
\frac{P^3x_1x_2 +P^2x_1x_3+P^1x_2x_3}{y_1\,y_2\,y_3}
+\frac{1}{2}\,\Bigg[y_1\,y_2\,y_3\,\sum_{i=1}^3 (P^i)^2y_i^{-2}   \nonumber \\
  && \hspace{-40pt} +\frac{1}{y_1\,y_2\,y_3}
  \,\Big(
   (P^3x_1x_2 + P^2x_1x_3+P^1x_2x_3)^2
   + \sum_{j\neq k\neq i=1}^3 (y_i)^2\,(P^kx_j+P^jx_k)^2\Big)\Bigg]
   \,.
\eeq
Inspired by \cite{Hotta:2007wz} we introduce the rescaled field  parameters and variables
($s_{ijk} = |\epsilon_{ijk}|$):
\begin{eqnarray}
  M_0^2 &=& 2\,\sqrt{-Q_0P^1P^2P^3}\,,\, R_i = \sqrt{-Q_0P^i\over {1\over 2}s_{ijk} P^jP^k}~,
   \label{eqn:newvar1}\\
  x_i &=&
  R_i\,t_i\,, \quad y_i =  R_i\,e^{\phi_i}~.
 \label{eqn:newvar2}
\end{eqnarray}
In terms of these, the black hole potential $V_{\text{BH}}$ simplifies to:
\begin{equation}
V_{BH}(t_i,\phi_i) = \frac{M_0^2}{4}\,e^{-\sum_r
\phi_r}\,\left(\sum_{i<j\,\,k\neq i,\,j} \left(e^{2(\phi_i+\phi_j)} +
(t_i+ t_j)^2\,e^{2\phi_k}\right) + (1+\sum_{i<j} t_i\,t_j)^2
\right)\,.
 \label{gvbh}
\end{equation}
The equations of motion are:
\begin{eqnarray}
  U^{\prime\prime} &=& e^{2U}\,V_{BH}\,, \label{Umotion} \\
  (t_i^\prime\,e^{-2\phi_i})^\prime &=& 2e^{2U}\,\frac{\partial V_{BH}}{\partial t_i}\,, \label{xmotion} \\
  \phi_i^{\prime\prime} + (t_i^\prime)^2\,e^{-2\phi_i} &=& 2e^{2U}\,\frac{\partial V_{BH}}{\partial \phi_i}\,,
  \label{ymotion}
\end{eqnarray}
and the constraint equation (\ref{eqn:constraint}) is:
\begin{equation}
  (U^\prime)^2 + \frac{1}{4}\sum_i \left[(\phi_i^\prime)^2 +
  (t_i^\prime)^2\,e^{-2\phi_i}\right] =
  e^{2U}\,V_{BH}\,. \label{consmotion}
\end{equation}

Yet another set of field variables $\{\beta,\,\alpha_i\}$ (i=1,2,3) will help us to disentangle
the coupled differential equations
(\ref{Umotion}-\ref{consmotion}):
\begin{eqnarray}
  \alpha_i = U + \frac{1}{2}\phi_i &  \Leftrightarrow & -2\phi_i = \beta + \sum_j \alpha_j - 4\alpha_i ~,
  \label{eqn:morenewvar1} \\
  \beta = U - \frac{1}{2}\sum_i \phi_i & \Leftrightarrow & 4U=\beta + \sum_j \alpha_j~.
  \label{eqn:morenewvar2}
\end{eqnarray}
The dynamical equations now become:
\begin{eqnarray}
  \alpha_i^{\prime\prime} + \frac{1}{2}(t_i^\prime)^2\,e^{-2\phi_i}
  &=& e^{2U}\,\left(V_{BH}+\frac{\partial
  V_{BH}}{\partial\phi_i}\right)\,, \label{eq1} \\
  \beta^{\prime\prime}- \frac{1}{2}\sum_i(t_i^\prime)^2\,e^{-2\phi_i}
  &=& e^{2U}\,\left(V_{BH}-\sum_i \frac{\partial
  V_{BH}}{\partial\phi_i}\right)\,, \label{eq2} \\
 \sum_i [(\alpha_i^\prime)^2-2\alpha_i^{\prime\prime}] +
 (\beta^\prime)^2 + \sum_{i<j} (\alpha_i^\prime-\alpha_j^\prime) &=&
 -2e^{2U}\,\left(V_{BH} + \sum_i \frac{\partial V_{BH}}{\partial
 \phi_i}\right)\,,
\label{eq4}
\end{eqnarray}
as well as (\ref{xmotion}). For the potential \eqref{gvbh} we have:
\begin{eqnarray}
  V_{BH} + \frac{\partial V_{BH}}{\partial \phi_k} &=&
  \frac{M_0^2}{2}\,e^{\phi_k -\phi_i-\phi_j}\,\left[(t_i+t_j)^2 +
  e^{2\phi_i}+e^{2\phi_j}\right]\,, \quad i,j\neq k~, \\ V_{BH}-\sum_k
  \frac{\partial V_{BH}}{\partial \phi_k} &=&
  \frac{M_0^2}{2}\,e^{-\sum_i \phi_i}\,\left[\sum_{i,j\neq k}
  e^{2\phi_r}\,(t_i+t_j)^2 + 2(1+\sum_{i<j} t_i\,t_j)^2\right]\,, \\
  V_{BH}+\sum_k \frac{\partial V_{BH}}{\partial \phi_k} &=&
  -\frac{M_0^2}{2}\,e^{-\sum_i\phi_i}\,\left[(1+\sum_{i<j} t_i\,t_j)^2
  - \sum_{i<j} e^{2(\phi_i+\phi_j)}\right]\,.
\end{eqnarray}
The dynamical equations still appear very complicated at this point
but they are in fact integrable.  It would be interesting to explore
the structure that makes this possible but we will be content with
simply finding the solutions, proceeding as
follows\cite{Hotta:2007wz}.  Suppose we could find solutions
$\alpha_i$ satisfying:
\begin{eqnarray}
  \alpha_i^{\prime\prime} &=&
  \frac{M_0^2}{2}\,e^{2\alpha_i}\,\left[e^{2(\alpha_k-\alpha_j)}+
  e^{2(\alpha_j-\alpha_k)}\right]~,\quad i\neq j \neq k~, \nonumber \\
  \sum_i [(\alpha_i^\prime)^2-2\alpha_i^{\prime\prime}] + \sum_{i<j}
  (\alpha_i^\prime-\alpha_j^\prime) &=& -M_0^2\,\sum_{i\neq j \neq k}
  e^{2(\alpha_j+\alpha_k-\alpha_i)}\,.
  \label{eq7}
\end{eqnarray}
If this is possible then \eqref{eq1} reduces to:
\begin{equation}
  t_i^\prime = \pm M_0\,(t_j+t_k)\,e^{3\alpha_i-\alpha_j-\alpha_k}\,, \quad j\neq k\neq i~,
 \label{eq5}
\end{equation}
and \eqref{eq4} reduces to:
\begin{equation}
  \beta^\prime = \pm M_0\,e^\beta\,(1+\sum_{i<j}t_i\,t_j)\,.
 \label{eq6}
\end{equation}
One can check that \eqref{eq5} and \eqref{eq6}
ensure that the remaining equations \eqref{eq2} and (\ref{xmotion}) are satisfied,
provided we take the same sign in both equations.
It is therefore sufficient to solve (\ref{eq7}-\ref{eq6}).

The first equation in \eqref{eq7} can be reorganized as:
\begin{equation}
 \left(\alpha_1 + \alpha_2 - \alpha_3\right)^{\prime\prime} =
 M^2_0 e^{2(\alpha_1 + \alpha_2 - \alpha_3)}\,,~~~({\rm and~cyclic~permutations})~,
\end{equation}
which is readily integrated as:
\begin{equation}
   e^{-(\alpha_j + \alpha_k - \alpha_i)} = a_i + M_0\tau\equiv \hat h_i ~,~~~({\rm and~cyclic~permutations})~,
  \label{sol1}
 \end{equation}
where $a_i$ is a positive integration constant.
One can check that this solution automatically satisfies the second equation in \eqref{eq7}.

Inserting the solution \eqref{sol1}  for $\alpha_i$ in the moduli equations \eqref{eq5} we find:
\begin{equation}
  t_i^\prime = \pm M_0\,(t_j + t_k)\,\frac{\hat h_i}{\hat h_j\,\hat h_k}~,~~~({\rm and~cyclic~permutations})~.
\end{equation}
It turns out that only the upper sign leads to regular solutions. For the lower sign
we have the solution:
\begin{equation}
  t_i = \frac{c}{\hat h_j\,\hat h_k}~,~~~({\rm and~cyclic~permutations})~.
\end{equation}
where $c$ is a new integration constant.

We are finally left to integrate (\ref{eq6}) which now takes the form:
\begin{equation}
  \beta^\prime = -M_0 e^\beta \left( 1 + \frac{c^2}{\hat h_1\, \hat
  h_2\, \hat h_3} \left(\frac{1}{\hat h_1} + \frac{1}{\hat h_2} +
  \frac{1}{\hat h_3}\right)\right)~.
\end{equation}
We find:
\begin{equation}
  e^{-\beta} = (a_0 + M_0\tau) - \frac{c^2}{\hat h_1\, \hat h_2\, \hat
  h_3} \equiv - \hat h_0 - \frac{c^2}{\hat h_1\, \hat h_2\, \hat h_3}~.
\end{equation}
Here $a_0$ is yet another integration constant which must be chosen sufficiently positive
so that $e^{-\beta}>0$ for all $0<\tau<\infty$.

We have thus found a solution in terms of four functions $\hat h_\Lambda$
($\Lambda=0,1,2,3$) that are linear in $\tau$ or, equivalently, harmonic on a spatial slice
of the black hole. There are five integration constants $c$,$a_\Lambda$.   We
transform to a more natural set of harmonic functions by asking that each one has a
$\tau$-dependence proportional to the corresponding dressed charge:
\be
\hat h_0 = -{M_0\over Q_0}\,H_0 ~,
\qquad \hat h_i = {M_0\over P^i}\,H^i ~.
\ee
If we use these new harmonic functions along with the changes of variables
(\ref{eqn:newvar1}-\ref{eqn:newvar2})
and (\ref{eqn:morenewvar1}-\ref{eqn:morenewvar2})
we can present the solution in terms of the physical warp factor:
\begin{equation}
  e^{-4U} = -\hat h_0\,\hat h_1\,\hat h_2\,\hat h_3 - c^2\, = 4H_0H^1H^2H^3 - c^2~,
\end{equation}
and the complex moduli fields:
\begin{equation}
  z_i = R_i \,\frac{c-i\,e^{-2U}}{{1\over 2} s_{ijk} \hat h_j\,\hat
  h_k}= \frac{c-i\,e^{-2U}}{s_{ijk} H^j\,H^k}~.
\end{equation}
We can now match the asymptotic conditions at infinity, $z^i \to B - i$ and
$e^{2U} \to 1$, by choosing $c = B $ and taking the other constants to be:
\be
a_0 = - {M_0\over \sqrt{2}\,Q_0}\, (1 + B^2)~,
\qquad a_i = {M_0\over \sqrt{2}\,P^i}~.
\ee
This completely specifies our seed solution.

\bibliographystyle{utphys}

\bibliography{GLS07}

\end{document}